\documentclass[conference]{IEEEtran}

\IEEEoverridecommandlockouts
\usepackage{cite}
\usepackage{amsmath,amssymb,amsfonts}
\usepackage{algorithmic}
\usepackage{graphicx}
\usepackage{textcomp}
\usepackage{xcolor}
\usepackage{booktabs}
\def\BibTeX{{\rm B\kern-.05em{\sc i\kern-.025em b}\kern-.08em
    T\kern-.1667em\lower.7ex\hbox{E}\kern-.125emX}}
    
\usepackage{multirow}
\usepackage{hyperref}
\usepackage{setspace}

\makeatletter
\def\@IEEEpubidpullup{8\baselineskip}
\makeatother

\usepackage{fancyhdr}
\usepackage{kantlipsum}
\fancypagestyle{plain}{
\fancyhf{}
\fancyhead[C]{Conference on \LaTeX} %% C or L or R.
}
\usepackage{eso-pic}

\begin{document}
\title{PureConnect: A Localized Social Media System to Increase Awareness and Connectedness in Environmental Justice Communities}


\author{\IEEEauthorblockN{1\textsuperscript{st} Omar Hammad}
\IEEEauthorblockA{\textit{Information and Computer Science Dept.} \\
\textit{King Fahd University of Petroleum and Minerals}\\
Dhahran, Saudi Arabia \\
omarjh@kfupm.edu.sa}
\and
\IEEEauthorblockN{2\textsuperscript{nd} Md Rezwanur Rahman}
\IEEEauthorblockA{\textit{Computer Science Dept.} \\
\textit{University of Colorado, Boulder}\\
Boulder, USA \\
mdra7255@colorado.edu}
\and
\IEEEauthorblockN{3\textsuperscript{rd} Gopala Krishna Vasanth Kanugo}
\IEEEauthorblockA{\textit{Computer Science Dept.} \\
\textit{University of Colorado, Boulder}\\
Boulder, USA \\
gopala.kanugo@colorado.edu}
\and
\IEEEauthorblockN{4\textsuperscript{th} Nicholas Clements}
\IEEEauthorblockA{\textit{Mechanical Engineering Dept.} \\
\textit{University of Colorado, Boulder}\\
Boulder, USA \\
nicholas.clements@colorado.edu}
\and
\IEEEauthorblockN{5\textsuperscript{th} Shivakant Mishra}
\IEEEauthorblockA{\textit{Computer Science Dept.} \\
\textit{University of Colorado, Boulder}\\
Boulder, USA \\
mishras@cs.colorado.edu}
\and
\IEEEauthorblockN{6\textsuperscript{th} Shelly Miller}
\IEEEauthorblockA{\textit{Mechanical Engineering Dept.} \\
\textit{University of Colorado, Boulder}\\
Boulder, USA \\
shelly.miller@colorado.edu}
\and
\IEEEauthorblockN{7\textsuperscript{th} Esther Sullivan}
\IEEEauthorblockA{\textit{Sociology Dept.} \\
\textit{University of Colorado, Denver}\\
Denver, USA \\
esther.sullivan@ucdenver.edu}
}

% \author{Omar Hammad, Md. Rezwanur Rahman, Nicholas Clements, Shelly Miller, Shivakant Mishra and Esther Sullivan}

% \author{\IEEEauthorblockN{1\textsuperscript{st} Omar Hammad}
% \IEEEauthorblockA{\textit{Information and Computer Science Dept.} \\
% \textit{King Fahd University of Petroleum and Minerals}\\
% Dhahran, Saudi Arabia \\
% omarjh@kfupm.edu.sa}
% \and
% \IEEEauthorblockN{2\textsuperscript{nd} Md Rezwanur Rahman}
% \IEEEauthorblockA{\textit{Computer Science Dept.} \\
% \textit{University of Colorado, Boulder}\\
% Boulder, USA \\
% mdra7255@colorado.edu }
% \and
% \IEEEauthorblockN{3\textsuperscript{rd} Nicholas Clements}
% \IEEEauthorblockA{\textit{Mechanical Engineering Dept.} \\
% \textit{University of Colorado, Boulder}\\
% Boulder, USA \\
% nicholas.clements@colorado.edu}
% \and
% \IEEEauthorblockN{4\textsuperscript{th} Shivakant Mishra}
% \IEEEauthorblockA{\textit{Computer Science Dept.} \\
% \textit{University of Colorado, Boulder}\\
% Boulder, USA \\
% mishras@cs.colorado.edu}
% \and
% \IEEEauthorblockN{5\textsuperscript{th} Shelly Miller}
% \IEEEauthorblockA{\textit{Mechanical Engineering Dept.} \\
% \textit{University of Colorado, Boulder}\\
% Boulder, USA \\
% shelly.miller@colorado.edu}
% \and
% \IEEEauthorblockN{6\textsuperscript{th} Esther Sullivan}
% \IEEEauthorblockA{\textit{Sociology Dept.} \\
% \textit{University of Colorado, Denver}\\
% Denver, USA \\
% esther.sullivan@ucdenver.edu}
% }

\maketitle
\IEEEpeerreviewmaketitle

\begin{abstract}

Frequent disruptions like highway constructions are common now-a-days, often impacting environmental justice communities (communities with low socio-economic status with disproportionately high and adverse human health and environmental effects) that live nearby. Based on our interactions via focus groups with the members of four environmental justice communities impacted by a major highway construction, a common concern is a sense of uncertainty about project activities and loss of social connectedness, leading to increased stress, depression, anxiety and diminished wellbeing. This paper addresses this concern by developing a localized social media system called PureConnect with a goal to raise the level of awareness about the project and increase social connectedness among the community members. PureConnect has been
designed using active engagement with four environmental justice communities affected by a major highway construction. It has been deployed in the real world among the members of the four environmental justice communities, and a detailed analysis of the data collected from this deployment as well as surveys show that PureConnect is potentially useful in improving community members' wellbeing and the members appreciate the functionalities it provides.

\end{abstract}


%    <concept>
%        <concept_id>10003120.10003138.10003140</concept_id>
%        <concept_desc>Human-centered computing~Ubiquitous and mobile computing systems and tools</concept_desc>
%        <concept_significance>500</concept_significance>
%        </concept>
%    <concept>
%        <concept_id>10003456.10010927.10003618</concept_id>
%        <concept_desc>Social and professional topics~Geographic characteristics</concept_desc>
%        <concept_significance>500</concept_significance>
%        </concept>
%  </ccs2012>

% \ccsdesc[500]{Human-centered computing~Ubiquitous and mobile computing systems and tools}
% \ccsdesc[500]{Social and professional topics~Geographic characteristics}


%%
%% This command processes the author affiliation and title

\section{introduction}

\par A major construction in North Denver, Colorado, USA took place from 2019 to 2022 to reconstruct a 10-mile stretch of the interstate highway (C70 project). The benefits, as stated on the project website are to increase safety, improve infrastructure, increase economic vitality, and many more \cite{cdot_central_2019}. However, this construction also caused increased noise and air pollution in the vicinity as well as increased stress in the communities living in the vicinity. The overall goal of our Social Justice and Environmental Quality - Denver (SJEQ) project ({\it https://www.sjeqdenver.com/}) is to partner with the members of the communities living the vicinity of C70 project (Globeville, Elyria-Swansea, Cole, and Clayton neighborhoods) to understand and address the disruptions caused and related neighborhood redevelopment. These four communities are  Environmental Justice Communities (EJC)---low socio-economic status with disproportionately high and adverse human health
and environmental effects\cite{bullard_environmental_1993}. The goal of SJEQ project is to first understand the negative impact of C70 on the health and well-being of these communities and then develop interventions to mitigate the negative impacts. Our project equipped and trained community members to use personal environmental sensors to monitor air pollution, and designed a set of apps where they can report their daily experiences related to construction and pollution.

\par To learn firsthand the community’s local knowledge
and concerns regarding the C70 project, we organized three focus groups in the Summer of 2021 with the residents of the four communities to discuss the topics of C70 construction and community concerns. Overall, 32 residents from these communities participated in our focus groups which included both English
and Spanish speakers. We obtained approval from IRB before conducting these focus groups. In addition, we have developed a smartphone app called PurEmotion to understand the impact of the C70 construction project on
people’s well-being (See https://www.sjeqdenver.com/ and Section \ref{sec:study} for more information on PurEmotion).

A major issue that we identified from the three focus groups (32 participants) and two deployments of PurEmotion (80-85 participants per deployment) was getting unclear and often misleading information about the construction. Of the residents who had the information, they had heard about the project updates through flyers left on their fences by the city and word-of-mouth from friends and family. Furthermore, due to frequent road closures and air/noise pollution, there was a feeling of loss of social connectedness in the community. Participants mentioned that they would like to have better distribution of information about road closures and alerts. They also requested resources and activities to help them manage their stress caused by a sense of loss of social connectedness.

\par In this paper, we describe the design, implementation and evaluation of a localized social media service called PureConnect. The goal of PureConnect is to increase awareness about the construction activities and improve social connectedness among the community members. PureConnect has been implemented as a slack workspace (https://www.slack.com) with several channels that community members may use to converse with other community members.
We aim to answer the following research questions: 

\begin{enumerate}

    \item How do we build a helpful software system to increase awareness and improve social connectedness among the environmental justice community members affected by a large construction project in their vicinity?

    \item What can we learn about environmental justice communities' behavior while using a localized social media service? 
    
    \item To what extent do the proposed intervention mechanism help in enhancing the well-being of environmental justice communities who live in the affected areas?
    
\end{enumerate}

PureConnect has been implemented as a Slack workspace that allows community members to connect with one another under our team's supervision. In addition,  relevant information about the construction project is posted by relevant authorities on this workspace. We deployed PureConnect in the four environmental justice communities over two different time periods (called cohorts), each lasting six to eight weeks. These deployments of PureConnect was done in conjunction with the deployment of the PurEmotion app. Overall, sixty-nine community members used PureConnect along with the PurEmotion app in these two cohorts. 

Feedback from the participants was overall positive, especially in the second cohort. In the first cohort, we have dedicated multiple channels for participants to communicate. However this caused confusion among the participants in deciding which channel is the right channel for the message they plan to post. Therefore, in the second cohort, we reduced the number of channels, resulting in better outcome. PureConnect includes use of different  Slack Apps (Apps that run within Slack such as an auto-translator) to assist in managing the workspace, integration of third party APIs to provide relevant value to our users and multiple engagement tactics to see what would trigger community member's engagement. Feedback from the participants after using PureConnect was quite positive and evaluation from the data collected from the deployment (PureConnect and PurEmotion) shows that there was an improvement in the well-being of people when they used PureConnect. In particular, this paper makes the following important contributions: 

\begin{enumerate}
    \item PureConnect is the first localized social media service to the best of our knowledge, specifically designed to address the needs of environmental justice communities facing disruptions due to major construction activities in their vicinity. 
    
    \item PureConnect has been built and experimented with using active engagement with the environmental justice communities affected by a major construction. 
    
    \item Multiple investigations in terms of real-world deployment and surveys show that environmental justice community users appreciate PureConnect functionalities and this service is potentially useful in improving their well-being. 
    
\end{enumerate}

%%%%%%%%%%%%%%%%%%%%%%%%%%%% Lit Review %%%%%%%%%%%%%%%%%%%%%%%%%%%%

\section{Literature Review}

\par Social ties and connection among community members are so important in times of hardship, for instance, to stand in the face of communities threats like crime and disasters \cite{sampson_great_2012}, or against large construction projects that would affect the nearby areas \cite{granovetter_strength_1973}. An important characteristic of a strong community is its social capital, which is the combined value of social structure and engagement of community members \cite{bourdieu_15_2008,coleman_social_1988,lin_building_1999}. High social capital can benefit communities during these hard times \cite{masden_tensions_2014} and it has been proven to enhance community members' attachment \cite{sampson_local_1988}, mental health \cite{elliott_stress_2000,ross_neighborhood_2000}, empowerment \cite{geis_new_1998} and trust \cite{ross_neighborhood_2000}. This is especially important in the US where civic engagement is becoming weaker \cite{putnam_bowling_2000}, and in places close to industrial and transportation areas that are subject to large construction projects that cause gentrification of low socio-economic people. 

\par Gentrification consists of combined steps of decisions and actions by higher-income people in areas where lower-income people live that would cause them to be displaced by force from their residence \cite{atkinson_does_2002,brown-saracino_gentrification_2013,lees_gentrification_2010,slater_eviction_2006}. Another type of gentrification could be voluntarily done by some members of the community due to some changes in the structural and cultural aspects of their community \cite{corbett_engaging_2019} due to the consequences caused by large projects. A big factor in such actions that none can overlook is race \cite{corbett_engaging_2019}, especially in a country with a long history of racial issues \cite{bonam_polluting_2016,keating_atlanta_2001}. 

\par All of this, and the decrease in social relations  has increased the interest of HCI researchers to study community connectedness and how can it stand in the face of such disruptions \cite{masden_tensions_2014,corbett_engaging_2019,lopez_behind_2017}. One of the major topics that have gained a lot of focus is social media and its role in either strengthening social capital \cite{masden_tensions_2014} or being against it \cite{corbett_engaging_2019}. In general, social media has been found to enhance social capital \cite{rainie_networked_2012} through the different ways that its users engage, like sharing photos, posts, and comments \cite{burke_social_2011}. 

\par When users of these systems are geographically close, these systems are called hyper-local \cite{lopez_behind_2017}, or community-focused social media \cite{masden_tensions_2014}. These local systems have a history of development in different forms such as bulletin boards \cite{colstad_community_1976}, forums \cite{rogers_pen_1994,dahlberg_extending_2001}, email lists \cite{carroll_developing_1996,hampton_neighboring_2003}, and later social networks for neighborhoods, like Nextdoor.com, online conversation communities, and marketplaces \cite{lopez_behind_2017}. Previous research has shown that these hyper-local systems contribute to helping communities in enhancing their social capital and connectedness \cite{oneil_assessing_2002}. A very relevant example of the benefit of such systems in strengthening communities is when  Toronto residents rallied and took action against a construction company that built their homes using internet-based tech \cite{bourdieu_15_2008}.

\par A well-known current local social media system that connects people who live in the same neighborhood together is Nextdoor.com \cite{masden_tensions_2014}. The system requires its users to use their real names as it is aligned with their goal “to create a safe, trusted environment where neighbors can connect with each other.” The system assigns a "lead" role to people who join a neighborhood first. These leads are giving permission to verify other joiners, moderate posts and adjust neighborhood boundaries. In order to allow new people to join a community, Nextdoor verifies them using physical mail addresses, or using their credit card information. In a semi-structured interviews study by \cite{masden_tensions_2014} on Nextdoor's users, they have found that many of the users already know each other before and they have chosen to join the system despite already using other social media apps \cite{masden_tensions_2014} which show the need of such systems. Many reasons were described such as the level of granularity that it had for each community, its affordance for conversations that its user's interface had \cite{masden_tensions_2014}. However, they have noticed that most people use nextdoor for business instead of social communication, some users's described the fixed split between neighbourhoods as limiting and not convenient in some cases, and that people were concerned about their privacy of data.

\par Despite the huge benefit of local social media systems on communities, there are some arguments that such systems can have negative impacts of neighborhoods, such as gentrification \cite{corbett_engaging_2019}. Corbet and Loukissas \cite{corbett_engaging_2019} studied how systems like Yelp, Nextdoor, and Zillow can mediate the process of gentrification. After analyzing the work of Sharon Zukin \cite{zukin_omnivores_2017} on the reviews of restaurants in New York in two neighborhoods, an African American and a Polish neighborhood on Yelp, Corbet and Loukissas \cite{corbett_engaging_2019} noted that some reviews mentioned the surrounding neighborhood and praised the gentrification process that is going on when talking about the African American restaurants, unlike the polish neighborhood where reviews were against gentrification. And since gentrification is a cooperative, and collective process, such reviews support or stand in the face of gentrification. Another system that was analyzed and claimed to mediate the process of gentrification is Nextdoor \cite{corbett_engaging_2019}. The system segregates neighborhoods into "Polygons" \cite{payne_welcome_2017} and allows each user to be part of one neighborhood. These boundaries can produce some social effects like spacial-economic fragmentation which re-inforces "existing class and racial boundaries in increasingly divided cities, drawing lines between ‘us’ and ‘them’ and amplifying the voices of neighbors who want to use the site to profile people they consider outsiders, even those who may have lived the area for generations” \cite{payne_welcome_2017}. Eventually, this results in gentrification \cite{corbett_engaging_2019}. In the mentioned systems' support for gentrification, it is important to notice that the negative effects are caused by outsiders to each community, not by people within. In fact, the majority of the interactions and communications that took place within the community members were positive to them. 

On top of their relative low-technical background, and varying ages, being a member in Environmental Justice (EJ) communities and undergoing continuous disruptions from construction projects gives its members specific traits such as mis-trust and loneliness. Although many researchers have investigated the impacts of deploying local social media systems \cite{bourdieu_15_2008,masden_tensions_2014,corbett_engaging_2019,rainie_networked_2012,burke_social_2011,oneil_assessing_2002}, none of them examined deploying those systems among EJ communities in particular.

%%%%%%%%%%%%%%%%%%%%%%%%%%%% Study Design %%%%%%%%%%%%%%%%%%%%%%%%%%%%

\section{Study Design}
\label{sec:study}

\subsection{Pre and Post intervention}
\label{sec:study-preintervention}

\par PureConnect is part of a larger project (SJEQ-Denver project) that was started in January 2022. The overall goal of SJEQ-Denver is to understand and mitigate the negative impacts of the C70 construction project on people's well-being. This project is structured over two phases, pre-intervention and post-intervention phases. The goal of the pre-intervention phase is to understand the impact of C70 construction on the health and wellbeing of the community members. This phase was comprised of deployment of the PurEmotion app over two different time periods of six-eight weeks called Cohort 1 and Cohort 2 
(January to July 2022) to collect wellbeing, location and air quality information from participants. PurEmotion is a smartphone diary app that community members use once a day to answer a few survey questions about their current feelings, their perceptions about the air quality around them and recent experience about their daily commutes.

\par Based on our findings from Cohort 1 and 2, we introduced three interventions: (1) PureConnect service to increase awareness about the construction project and improve social connectednesss, (2) indoor air cleaners to improve the indoor air quality, and (3) PureNav service to help community members navigate easier \cite{hammad_purenav_2023}. All these interventions were deployed together over two different time periods (Cohort 3 and Cohort 4) of six-eight weeks each. PurEmotion was deployed in these cohorts as well to assess the effectiveness of these interventions. Details of PureNav design, implementation and evaluation are described in \cite{hammad_purenav_2023}, while information about the air cleaners is available at
{\it https://www.sjeqdenver.com/}. This paper focuses on PureConnect.

\subsection{PureConnect}

\begin{figure*}
    \centering
    \includegraphics[width=\textwidth]{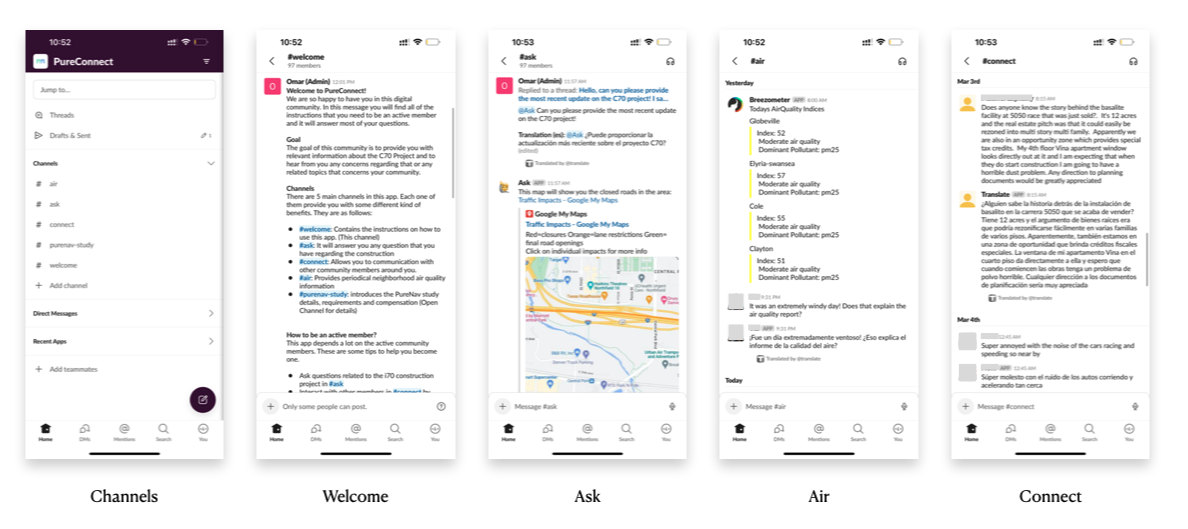}
    \caption{The main screens of Pureconnect app}
    \label{fig:pureconnect_ss}
\end{figure*}

\subsubsection{System Design}

%\par We learned from the first phase of the study that two of the most common issues that community members faced were community belonging and unclear information about the C70 construction project. In particular, many participants feel that they were not listened to by the city officials while making decisions about the construction project although the project has a direct effect on them. Moreover, they felt that the distribution of information about the construction project was not done properly. For instance, they hear about the updates of the project from friends and family and word of mouth. 

\par To address the problems of a sense of uncertainty about the project activities
and loss of social connectedness caused by the construction activities, we have designed and evaluated a loaclized social media service called PureConnect (See Figure \ref{fig:pureconnect_ss}). Our initial set of requirements for this service are to include the following functionalities:

\begin{itemize}
    \item A Q/A interface to answer community members' common questions about the construction project.
    \item A way for the community members to be able to express their concerns about the issues that they encounter.
    \item A forum to allow community members to discuss relevant community issues related to the C70 project.
\end{itemize}

\par After analyzing the requirements of this service we decided to use a Slack workspace (www.slack.com) and build these individual Slack apps to provide the required functionalities. Slack is a messaging system that was designed initially for businesses, however, a lot of different interest groups use it for communication and discussions \cite{chatterjee_exploratory_2019}. Slack is a suitable platform for many reasons, such as reliability, its support for plugins, easy APIs and it is free for general use. In addition Slack allows us to deliver functionality without the need to ask users for continuous updates, like what happens with traditional apps. However, there are shortcomings of using slack such as limited UI options and the trial version will hide messages after 30 days.

\par After deciding on Slack we needed to design the workspace to be aligned with our requirements. In the first version which we deployed in Cohort 3, we created the following slack channels:

\begin{itemize}
    \item \#welcome: on-boarding and usage instructions.
    \item \#ask: chat-bot to answer questions about the construction.
    \item \#express: express concerns about the project.
    \item \#suggest: suggest ideas to help mitigate issues.
    \item \#connect: allow community members to communicate with each other however they want.
    \item \#air: periodic neighborhood air quality information. 
    \item \#info: periodic information about the current activities and status of the construction project.
    \item \#researchers: a private channel for researchers to discuss any issues.
    \item \#ask\_research: a private channel for admins to discuss any issues.
\end{itemize}

\par Based on the feedback we received from the community members who used our service in the first deployment, we noticed that people were confused about the purpose of three channels, \#express, \#suggest and \#connect channels and suggested to keep a single channel instead. So in  the revised version of the service that we deployed in Cohort 4, we kept a single channel \#connect, and removed the other two.

\subsubsection{Onboarding}

\par We posted and pinned a welcoming message in the welcome channel (Figure \ref{fig:pureconnect_ss}, Welcome) to define the rules and guidelines of using the workspace. The welcoming message included a general statement that welcomed people to the channel, the goal of the workspace, a description of the channels, a guide on how to be an active member, posting guidelines, profile setup instructions, help and information regarding Spanish translation.

\par This is how we defined our goal to the users:
\begin{quote}
\textit{The goal of this community is to provide you with all the information that you need to know about the Central 70 Project and to hear from you any concerns regarding that. }
\end{quote}

\par The posting guidelines were inspired by online discussion guidelines and included the following \cite{goldberg_7_2020}: 

\begin{itemize}
    \item Be open: Feel free to express your opinions without fear of being judged
    \item Be a good listener: Read previous posts before you post a new one
    \item Be positive: remain positive in your posts and help others
    \item Be brief: concise and clear posts attract more engagement
    \item Be respectful: when you disagree respect other peoples’ opinions
    \item Be focused: Keep discussion related to your neighborhood experiences
\end{itemize}

\par With user-generated content systems comes the risk of misuse. We have dealt with this by including these guidelines and assigning moderators from the research group. To make sure that people get help when they need, we developed a simple slack app called Help that users can trigger by typing @Help that sends a message to a private channel called \#ask\_research that has the research team only so that they can enable notifications for this channel and respond to it in a timely manner. 

\par Since a large number of community members are Spanish speakers, we integrated a Slack app called Translate (https://apps.mt-hacks.com/translate/). Translate is a third-party application that allows language detection and translation. The app is installed on the PureConnect workspace and configurable by interacting with the application under the app's section.  The current configuration enables detection and translation for the primary languages that are English and Spanish. 

\subsubsection{Dialogflow Chatbot Integration}

\par The \#ask channel was integrated with Google's Dialogflow. Dialogflow is "a natural language understanding platform from Google that helps developers to design conversational interfaces and integrate them into their applications." \cite{sabharwal_introduction_2020}. We compiled a set of commonly asked questions about the C70 construction project from C70 website \cite{cdot_central_2019} as well as the questions we heard from the community members in the three focus groups and the two phases of deployment of PurEmotion app (Pre-intervention study in Cohorts 1 and 2). Dialogflow allows a chatbot developer to define a set of \textit{intents} which is "the verb or action that is part of the conversation with the user" \cite{sabharwal_introduction_2020}. For instance "what is your name?", "what's your name?" have the same intent. We defined the following intents: 

\begin{itemize}
    \item How can I report road closures?
    \item How can we communicate our concerns?
    \item How is it impacting the air quality? 
    \item How will environmental concerns be addressed?
    \item How will the construction project help the community?
    \item How would the research project help?
    \item I want to know more about the construction project.
    \item What can I do to be safe from air pollution?
    \item What is the status of the construction?
    \item What roads will be closed due to the construction project?
    \item Which neighborhoods are affected by this?
    \item Who is involved in the construction?
\end{itemize}

\par For each of intent, we researched and provided an answer to the best of our ability mainly from the project website. We built another instance of the chatbot with the same intents in Spanish language as well. There was no easy way to do a multi-language agent and so we translated each set of questions and answers, and verified with a native Spanish speaker to ensure that the translation was accurate.

\par To allow users to interact with the chatbot, we have developed a Slack app for each language that can be triggered using the keyword @Ask for the English app and @Pregunta for the Spanish app. People were also able to chat with the chatbot in direct messages that are not visible to anyone in the community. In case a question that was asked was not part of any intent, that question was stored in the Dialogflow panel so that developers could provide an answer and update the set of of intents.

\subsubsection{COTrip API Integration}

\par The \#info channel was linked to a cloud function that calls an API from the Colorado Department of Transportation (https://www.codot.gov/) that provides information about planned construction events. The API is part of a set of APIs provided by (cotrip.org) which provides periodic updates about road activities that happen in the State of Colorado. To run the function on a schedule, we developed an AWS Lambda function that runs each day at 9:00 AM and calls the \textit{Planned Events} API. The API does not support the filtering of data so we filtered the data after receiving all the data points from the API (max 1000) data points. The filtering function takes into account the latitude and longitude of each reported event and filters out any points that are farther than 10 km from the middle point of the four communities of interest. The function only reports the events that occur on the day of reporting. 

\subsubsection{Engagement Plan}

\par To remain active and engaging, online groups should have valuable content that benefits the members \cite{lopez_behind_2017}. In PureConnect, this content is expected to be generated by the participants themselves after a while, however, there has to be a trigger initially to motivate the participants to share content. In our first cohort (Cohort 3), we did that by preparing a content plan that had some useful information that we shared on a regular schedule. Along with that, we have encouraged the research group the following: 

% interactions numbers table
\begin{table*}
    \centering
    \caption{Interactions Numbers}
    \begin{tabular}{lllllll}
    \toprule
    {} &  Reactions &  Messages &  \# Users &  Reactions / User &  Messages / User &  Reactions / Messages \\
    cohort &            &           &                  &                     &                    &                        \\
    \midrule
    3   &      94 &     39 &            14 &             6.71 &            2.78 &                   2.4 \\
    4      &        261 &       358 &               55 &                4.74 &               6.50 &                   0.72 \\
    \bottomrule
    \\
    \end{tabular}
    \label{tab:interactions-numbers}
\end{table*}

\begin{itemize}
    \item Mutual Support: Provide mutual support to participants so they can help each other out.
    \item Post High-Quality Content: Provide content of high quality to keep participants engaged.
    \item Storytelling: Use storytelling as a way to engage participants and build relationships.
    \item Celebrate Accomplishments: Recognize and celebrate participants’ accomplishments to motivate.
\end{itemize}

\par The plan included a daily schedule of a mix of useful resources, updates to the channels, new features, polls and questions and discussion points. The following are examples of some resources shared with them:

\begin{itemize}
    \item https://maps.cotrip.org/ (A map that shows road incidents)
    \item https://twitter.com/ColoradoDOT
    \item https://www.waze.com/
    \item https://www.breezometer.com/air-quality-map/air-quality
\end{itemize}

\par In addition we shared some preliminary results from the PurEmotion study like a graph showing correlation between bad environmental quality and the level of happiness, or a graph showing correlation between how close people are to construction activities and their level of happiness. With each figure, we asked them if they agree or disagree with the preliminary results. 

% \subsubsection{System architecture}
% Figure \ref{fig:pureconnect_backend} shows the system architecture of Pureconnect. In the front end we had users interact with the following apps: Help bot, ask bot, slack reader bot, translate bot, Breezometer bot and c70 activity lambda. All of these apps communicated with a slack server to fetch or send their data. Two of the apps were hosted on their own 3rd party servers while two were hosted by the research team on Amazon AWS and Google Cloud. 

% % % % % % % % % % % % % % % % % % % % % % % % % % % %  Analysis % % % % % % % % % % % % % % % % % % % % % % % % % % 

\section{Evaluation}

\par PureConnect Version 1 was deployed among 14 users in Cohort 3 (October - December 2022) and Version 2 (with reduced number of channels) was deployed among 55 users in Cohort 4 (March - April 2023). To evaluate the efficacy of PureConnect, we analyze three aspects of data collected from the two deployments, level of user interactions via PureConnect, sentiments of the messages posted by users in PureConnect, and emotions of the PureConnect users.

\subsection{Users Interactions}

\par There are two types of user interactions in PureConnect: Messages posted by the users and Reactions provided by the users on posted comments. A reaction on Slack is when a user taps on a message and adds an emoji reaction. Table \ref{tab:interactions-numbers} shows the details of user interactions in Cohort 3 and 4. We see that overall, the number of interactions in Cohort 3 were about one fifth the number of interactions in the Cohort 4. This is partly because of the lower number of users in Cohort 3. However, we see that the average number interactions per user increased from 9.5 (Cohort 3) to 11.25 (Cohort 4). We attribute this increase in higher user interactions as well as our ability to recruit more user in Cohort 4 to increased awareness of PureConnect over time, which indicates that the users find PuerConnect useful. Not all of the participants were active throughout study, however, almost 10\% of them were highly active interacting more than 30 times over the two months of deployment period, which is a common usage pattern in interactive social media services.

We also note that user interaction was mainly done using reactions in Cohort 3 (6.71 reactions/user vs 2.78 messages/user), while in Cohort 4, user interaction was mainly done by posting messages (4.74 reaction/user vs 6.5 message/user). Recognizing that posting a message takes much more effort than reacting, this shows that users found the second version of PureConnect (reduced number of channels) much more useful and engaging.

\begin{table}[htbp]
  \centering
  \begin{minipage}{0.45\textwidth}
    \centering
    \caption{The top reactions used}
    \begin{tabular}{lr}
      \toprule
      Reaction & count \\
      \midrule
      +1 & 241 \\
      raised-hands & 19 \\
      heart & 15 \\
      flushed & 15 \\
      slightly-smiling-face & 10 \\
      wave & 9 \\
      heart-eyes & 9 \\
      open-mouth & 8 \\
      worried & 6 \\
      joy & 5 \\
      \bottomrule\\
    \end{tabular}
    
    \label{tab:emojies_counts}
  \end{minipage}
  \hfill
  \begin{minipage}{0.45\textwidth}
    \centering
    \caption{Average number of interactions per message for each channel}
    \begin{tabular}{lll}
      \toprule
      channels & \multicolumn{2}{l}{Avg Interaction / user} \\
      {} & Cohort 3 & Cohort 4 \\
      \midrule
      air & 1.71 & 4.5 \\
      connect & 2.8 & 4.5 \\
      ask & 3.33 & 3.03 \\
      express & 1.19 & N/A \\
      info & 0.48 & N/A \\
      suggest & 0.62 & N/A \\
      \bottomrule\\
    \end{tabular}
    
    \label{tab:channels_avgs}
  \end{minipage}
\end{table}

Table \ref{tab:emojies_counts} provides the details of the type of reaction users responded with. We see that the top emoji used is (Thumbs-up) \textbf{+1}. Further, we notice that most of the reactions were positive reactions and very few reactions were negative. This indicates that users generally felt positive while using the PureConnect service. Finally, Table \ref{tab:channels_avgs} shows the average number of interactions per user in each channel in the two cohorts. Note that the PureConnect version used in Cohort 4 had fewer channels compared to the number of channels in the first version used in Cohort 3. We can see that in Cohort 3, \#ask channel had the most average number of interactions per user. However, in Cohort 4, when we reduced the number of channels, the average number of interactions per user in the \#air and \#connect channels increased significantly, while the average number interactions per user in the \#ask channel remained about the same. Combining this with our earlier observation that the average number of interactions per user increased from Cohort 3 to Cohort 4, we can say that reducing the number of channels had a positive effect, possibly because people were less confused by which channel to post in.
  
\subsection{Sentiment Analysis}

To further evaluate the efficacy of PureConnect, we conducted Vader \cite{hutto_vader_2014} sentiment analysis overall, for each cohort separately and for each channel separately on the messages posted by the users. Vader is a lexicon-based sentiment analysis method based on more than 7,500 lexical features. It can handle informal text like social media content. In the overall analysis, we found that the sentiments were close to neutral in both cohorts, but on the positive side (around 0.15). Further, the difference in the sentiments between the two cohorts was not significant.

\subsubsection{Channel Sentiment}
\par
\begin{figure}
\begin{minipage}{0.45\textwidth}
    \centering
    \includegraphics[width=\textwidth]{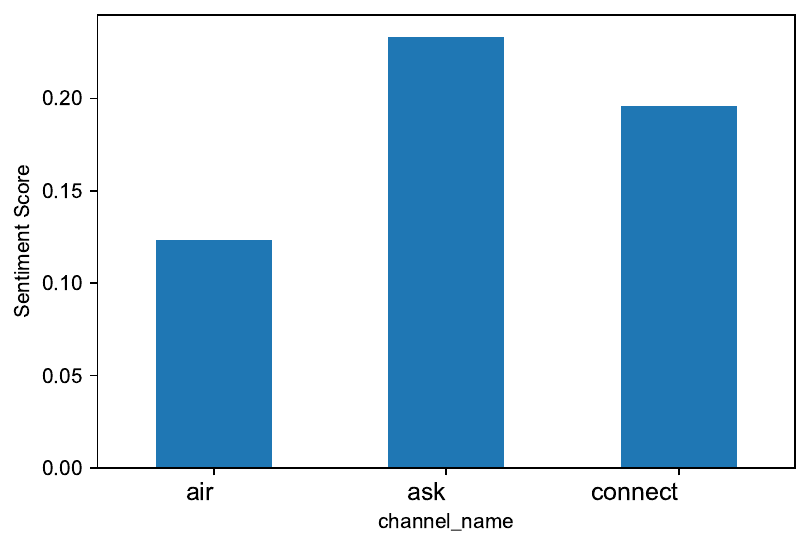}
    \caption{The average difference in sentiments among channels}
    \label{fig:sent_channel}
\end{minipage}
\hfill
    \begin{minipage}{0.45\textwidth}
    \centering
    \includegraphics[width=\textwidth]{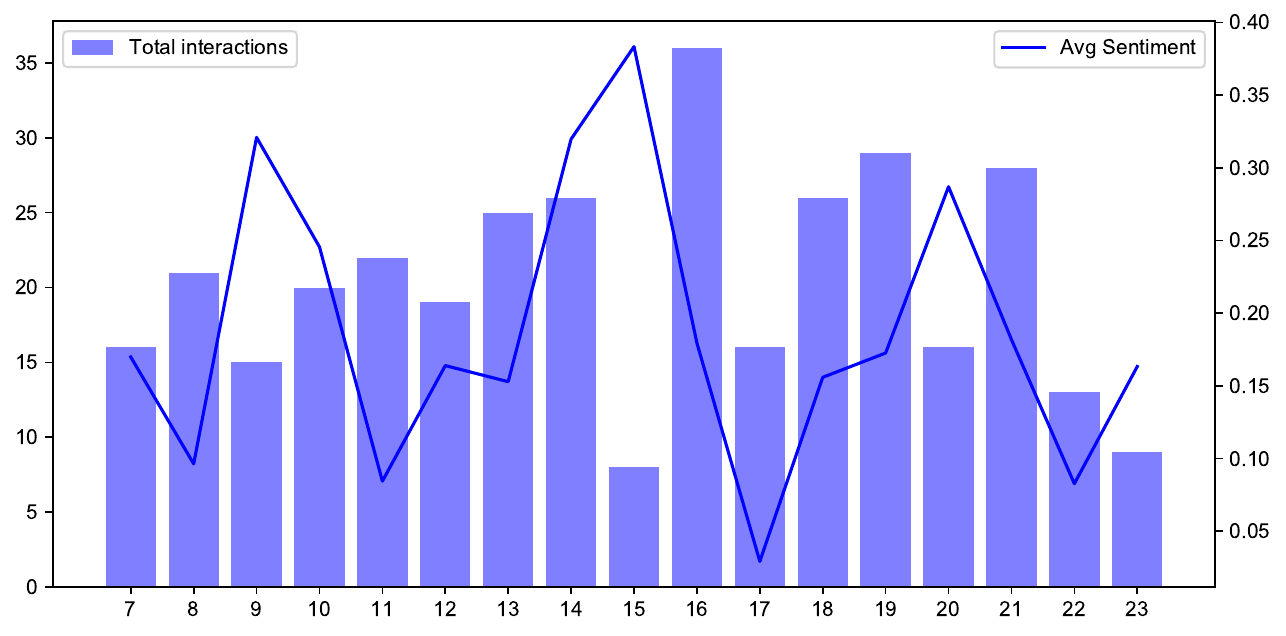}
    \caption{The average difference in sentiments over different 1-hour time periods (7 AM - 11 PM)}
    \label{fig:sent_hour}
\end{minipage}

\end{figure}

 Next we look at the differences in sentiments across different channels. Looking at all channels,
%(See Figure~\ref{fig:sent_hour}, 
we did not find any significant difference in sentiments, however we noted that some of the channels had only a few interactions. So, we looked closely at the three channels with most number of interactions, namely \#air, \#ask and \#connect. We found that the \#air channel had the least positive sentiment average (p-value = 0.01) followed by \#ask and \#connect that are close to each other (See Figure \ref{fig:sent_channel}). This indicate that users were concerned about the air quality and increased air pollution due to the construction.

\subsubsection{Time-related sentiments}
\par We observed that the sentiments of messages being posted changed during different times of the day. Looking at the average sentiments for each hour during the day from 7 AM to 11 PM (See  Figure \ref{fig:sent_hour}), we see that the least positive sentiment scores are at 5 PM and the most positive sentiment scores are at 3 PM. To understand this further, we investigated several possible reasons such as whether there are some people who posted only at some specific hours or the impact of the channel the users posted in. However, we couldn't verify any of these reasons. We also looked at other aspects like weeks, months, neighbourhoods and distance from the center of construction, but there was no clear relationship between the sentiments and these aspects. 

\subsubsection{Words Associations}
\par
\begin{figure}
    \begin{minipage}{0.45\textwidth}
    \centering
    \includegraphics[width=\textwidth]{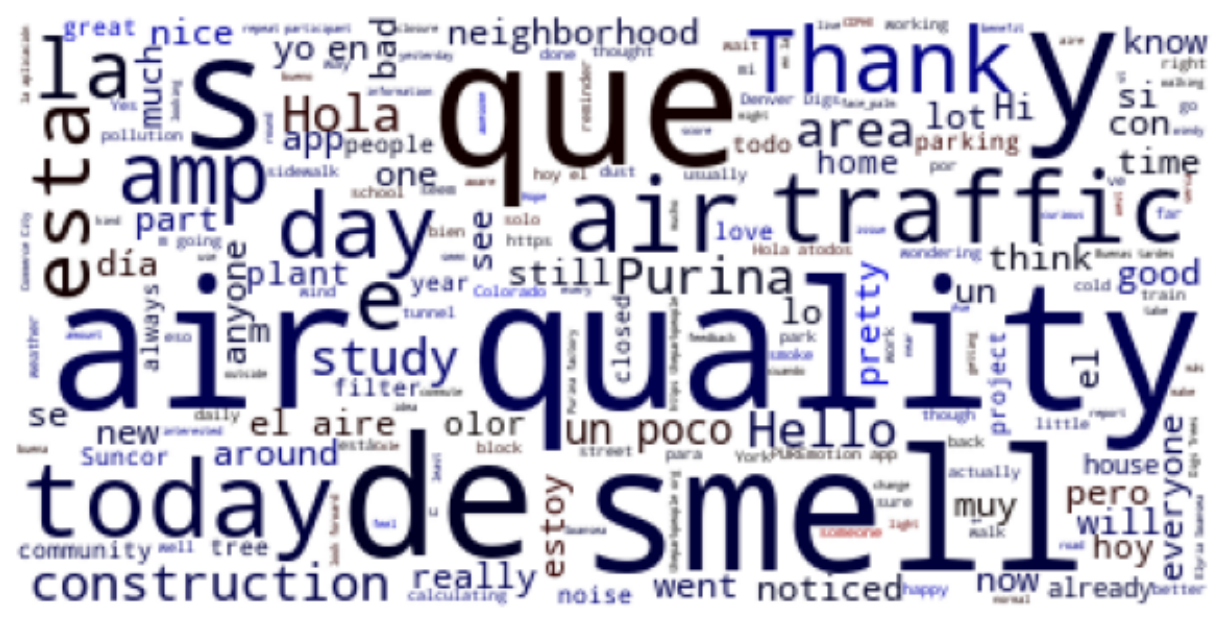}
    \caption{Word cloud for all messages}
    \label{fig:wordcloud_1}
\end{minipage}
\hfill
\begin{minipage}{0.45\textwidth}
    \centering
    \includegraphics[width=\textwidth]{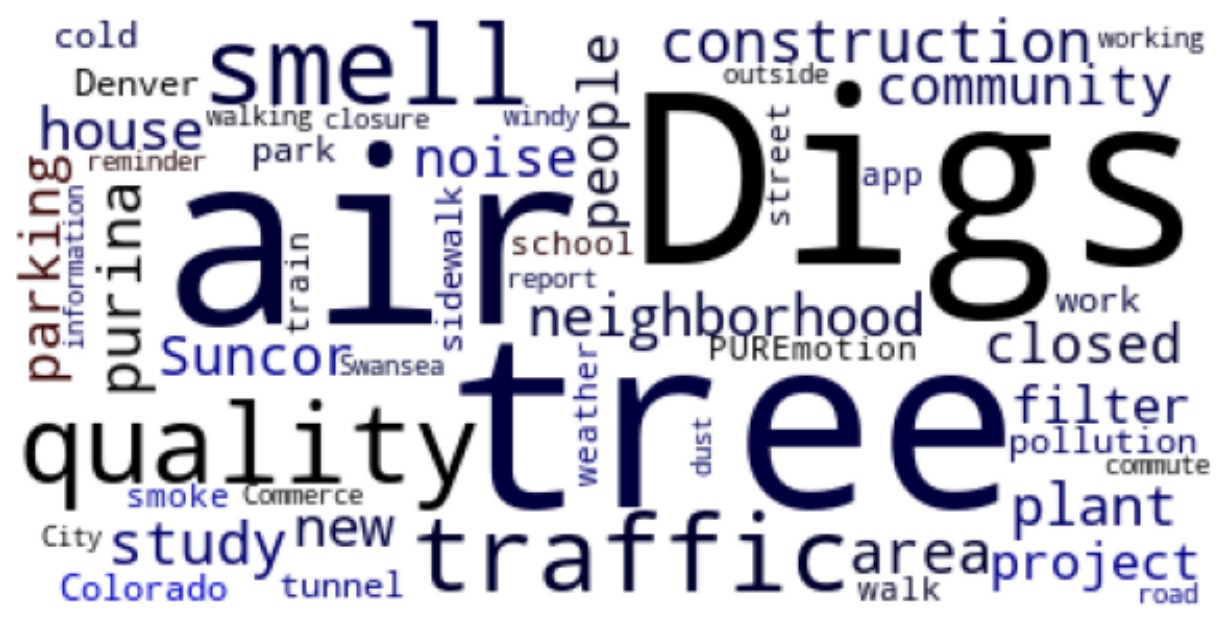}
    \caption{Word cloud for top relevant words messages}
    \label{fig:wordcloud_top}
\end{minipage}
\end{figure}

 Next we looked at the association between the words related to the study and the sentiment scores. We started by drawing a word cloud of all the words and the associated sentiment scores of the sentences that they belonged to. Figure \ref{fig:wordcloud_1} shows the most occurring words colored from red (negative) to blue (positive). Some of the relevant words are air, smell, traffic, construction, neighborhood, and Purina (a dog food factory nearby). We can see from the figure that most of the words are black which shows that they were neutral. This is because these words were contained in both positive and negative sentiments, although mostly positive. 

% words clouds figures

\par We then focused on a set of 50 words \ref{tab:top_words} that are most closely related to the project. We have done the same word cloud analysis and coloring on these words and the results are shown in Figure \ref{fig:wordcloud_top}. The most mentioned words (Large fonts) were air, digs, tree, traffic, smell, and quality. The words that were mostly with positive sentiments (Blue) were Colorado, Project, Suncor, noise, smoke and road. Although some of these words look like negative impacts on the community, however, people's messages were positive that these will be changed. For instance the following quote from a participant talked about some concerns but at the end showed a positive attitude that things will change positively:
\medskip

% top words table
\begin{table}[h]
\centering
\caption{Word Categorization Table}
\resizebox{\columnwidth}{!}{%
\begin{tabular}{|ccccc|}
\hline
Air quality & Smell & Traffic & Area & Purina \\
Construction & Study & Neighborhood & New & Plant \\
House & Community & Suncor & Tree & Closed \\
Noise & Parking & People & Project & Filter \\
Pollution & Park & Denver Digs & Colorado & Cold \\
Sidewalk & Work & Tunnel & PUREmotion app & Walk \\
Train & Smoke & Weather & School & Street \\
Working & Reminder & Dust & Swansea & Outside \\
Road & Information & Commute & Windy & Closure \\
Commerce City & Digs Trees & Walking & Report & \\ 
\hline
\end{tabular}%
}
\medskip
\label{tab:top_words}
\end{table} 

\begin{quote}
    \textit{Glad to be here as well!  I am concerned with air quality due to construction, and although it’s not a part of this study, I am also concerned by my proximity to the Suncor oil refinery.  I can see it clearly from my front porch, and I get a front row seat every time it malfunctions, catches fire, and spews pollutants into the air….  \textbf{Anyways!  I am passionate about air quality, community, and I am excited to be a part of this study}!}
\end{quote}

Another quote by a participant shows how their participation in the study would help in shutting down the Suncor oil refinery: 

\begin{quote}
   \textit{ Hi all, happy to be a part of the study.  \textbf{Hope we can get Suncor shut down!}}
\end{quote}

\par On the other hand, the words that were associated with negative sentiments were parking and school. People in their messages were mostly complaining when talking about parking. For instance, the following quote by a participant talked about some traffic issues and included parking also:

\begin{quote}
    \textit{same! i live on franklin and 35th where they’ve closed it down. \textbf{frustrating for my commute and lack of parking} for everyone who lives there (which already had a parking problem)}
\end{quote}

And here is another participant trying to notify other participants about some road problems that he faced:

\begin{quote}
    Warning! \textbf{ 36th \& Blake Street Closure by RTD}\textbf{ warning\_emoji} Spoke to a construction worker this morning \& he explained that the closure was temporary \& lanes/parking should be reopened by Monday.
\end{quote}

\par But in general, the words were mostly associated with positive sentiments, which tells us that people were in a positive, and hopeful situations while communicating through PureConnect. 

\subsection{PureConnect and Feelings Correlation}

\par As mentioned earlierin Section~\ref{sec:study-preintervention}, PureConnect users used PurEmotion app as well in which they answered a few survey questions at least once every day about their current
feelings, their perceptions about the air quality around
them and recent experience about their daily commute. One of the question in PurEmption app prompts users to give a score on the following feelings from 1 to 5: Happy, Distressed, Irritable, Alert/Awake, Lonely. We analyzed correlation between the frequency of usage of PureConnect by users (number of user interactions---messages or reactions) with their feelings. Conducting Kendall's correlation analysis gave medium correlation values as seen in Table \ref{tab:msg_corr}. We see that PureConnect usage is positively correlated with positive feelings and negatively correlated with negative feelings. To dig deeper we combined the low, medium and high levels of each feelings (low: 1 or 2; medium: 3; high: 4 or 5) and calculated the average number of interactions. We found that the average number of interactions was higher for users who have higher positive feelings (Happy). Similarly, and the average number of interactions was lower for users who have higher negative feelings (Distressed, Irritable, Lonely). For Alert/Awake, the pattern was not that clear. Conducting the ANOVA test for all feelings among the three level averages gave p value < 0.0001, which shows that there is a significant difference among them. Overall, this analysis shows that the usage of PureConnect helps the emotional state of the users positively. Note that this is based on correlation and causation.

\begin{table}[h]
\centering
\caption{Kendall Correlation values for different feelings with Number of submitted interactions in PureConnect}
\begin{tabular}{|c|c|}
\hline
\textbf{Feeling} & \textbf{Correlation with \#Interactions} \\
Happy & 0.199 \\
Distressed & -0.233 \\
Irritable & -0.223 \\
Alert/Awake & 0.140 \\
Lonely & -0.200 \\
\hline
\end{tabular}
\medskip
\label{tab:msg_corr}
\end{table}

\section{Intervention Questionnaire Survey} 

To further evaluate the usefulness of PureConnect, we conducted a survey among all PureConnect users at the end of each of the cohorts (Cohort 3 and Cohort 4). This survey included the following general questions:

\begin{enumerate}
    \item How helpful do you find PureConnect? (1 Not helpful at all, 10 Extremely helpful)
    \item How easy do you find it to use PureConnect? (1 Not easy at all, 10 Extremely easy)
    \item What did you like about PureConnect? (open ended)
    \item What did you dislike about PureConnect? (Open ended)
    \item Will you continue using the app after the study period? (Yes, No)
    \item How likely are you to recommend PureConnect to others? (1 Not at all - 10 Extremely)
    \item Do you have any suggestions or comments for future releases of PureConnect?
\end{enumerate}

\par In total we received 40 responses from the two cohorts, which is about 58\% of the total PureConnect users. We provided a compensation of \$5 to the users for completing the surveys. Overall, users reported mixed feedback on helpfulness in Cohort 3. Users liked being connected to others and receiving air quality and odor information. However, they reported problems with usage and lack of in-depth interactions with others. In Cohort 4, helpfulness increased by 34\% (from -0.1 to 0.2). This indicates that people appreciated having fewer chat channels, and more responsiveness to their messages.
Average easiness slightly increase in Cohort 4 (0.53 to 0.63) after reducing the number of channels and providing more instructions regarding the purpose of PureConnect. 

\par To learn how valuable PureConnect is compared to other social media apps, we asked the following questions:

\begin{itemize}
    \item With regard to C70 Construction, what \textit{other social media apps} have you used to connect with your community and discuss shared issues? (Nextdoor, Twitter, Facebook, Instagram, Reddit, Other)
    \item Compared to other social media apps that you have used, how valuable is PureConnect in \textbf{increasing your awareness} about C70? (1 Not valuable at all - 10 Extremely valuable )
    \item Compared to other social media apps that you have used, how valuable is PureConnect is \textbf{making you feel connected} with other community members? (1 Not valuable at all - 10 Extremely valuable )
    \item Compared to other social media apps that you have used, how valuable is PureConnect in \textbf{making you feel listened to} regarding C70? (1 Not valuable at all - 10 Extremely valuable )
\end{itemize}

\par Survey responses show that compared to other social media apps, users found PureConnect valuable in increasing their awareness about construction project activities (Avg 0.6 on a -1 to 1 scale), making them feel connected (Avg 0.4) and making them feel listened to (Avg 0.2). Overall, we can conclude from these results that PureConnect does help with increasing awareness about the construction activities and improving social connectedness among the community members, which was the goal of PureConnect.

\section{Discussion}

\par The goal of developing PureConnect was to contribute in mitigating the negative impacts of the C70 construction project on people's well-being. We began this study with three research questions.

\subsection{Building a helpful system to support awareness and connectedness}

\par Our first research question was: How do we build a software system to increase awareness and improve social connectedness among environmental justice community members affected by a large construction project in their vicinity? We have shown in the last section how users found the system helping them in increasing their levels of awareness about the construction project activities and connectedness among community members. 
%Awareness was the most clearly achieved goal from the surveys, while connectedness was somehow achieved. 

\par One supporting evidence for connectedness was the feeling of loneliness score that we measured using the PurEmotion app. That question asked users each day the following: Rate the level of loneliness from 1 to 5. We conducted two tests that showed that the level of loneliness decreased for users who used PureConnect. More clearly, we have shown that there is a high negative correlation between the number of interactions that users had in PureConnect and the feeling of loneliness. The more people used PureConnect, less lonely they felt (Kendall correlation of -2.0).

\par When we first deployed PureConnect over slack, we had assumed that the solution will be easy to use and straightforward since slack is a widely used tool. However, for a lot of our users, it was a first time experience. So it's important to know beforehand the level of technical background that the users have and build on top of that.

\par Another finding on how to build such a system was that some of the information that we were sharing in Cohort 3 was not very relevant to several users. When we focused on more relevant information in the Cohort 4, users reported higher helpfulness score for the system. So, it is important to understand the needs of the participants beforehand and provide relevant information accordingly. For instance, one of the most appreciated piece of information that we shared and was helpful for users was air quality since it was one of the major concerns of the community.

\par Furthermore, users did not like when they asked questions and no one responded to them with in a reasonable time period (1-2 days). The rate of responsiveness is important, because timely responsiveness makes people feel that they are being listened to. For instance one user highly appreciated when they asked about a construction work that was happening close to their house and one of the research team member followed up with city officials regarding the issue and reported back to that user.

\subsection{Local social media usage behaviour for environmental justice communities}

\par Our second research question was: What can we learn about environmental justice communities behavior while using a localized social media app? From what we observed, there are several things to learn from this study with regards to the usage behaviour of local social media systems.

\par First, it is important to keep the social media app simple with few functionalities and provide only the information users care about, otherwise app usage decreases. For instance in Cohort 3, we had a large number of channels and users could not decide which channel to post on. So we observed that people were mostly reactive in Cohort 3 where the number of reactions were significantly higher than posting messages. However, in Cohort 4 where we reduced the number of channels to three, we observed that users posted significantly more messages on average. 

\par Another clear observation that we had is the overall positive way users discuss issues in these settings. Although there were complaints about several issues like smell and traffic, users generally tried to present their thoughts positively. 
%we believe this is one of the effects of supervision by the research team.  

\par Finally we observed that user sentiments are affected by the time of the day and the scenario that they are in. For instance, we observed that for the \#air channel, the sentiment was lower than the \#ask channel. This is possibly because in the \#air channel, users are presented with air quality data which was generally poor and users could directly relate to. In addition, during different hours of the day, users' average sentiments were different. For example sentiments had lowest score at 12 PM and highest score at 2 PM. We hypothesize that this partly due to external factors such as being hungry at 12 PM.

\subsection{The effect of the intervention on users' well-being}

\par Our last research questions was: To what extent do the proposed intervention mechanism help in enhancing the well-being for environmental justice communities who live in the affected areas?

\par In order to answer this question we analyzed the data from PureConnect usage as well as two surveys that administered at the end of each cohort. Based on this analysis, we identify four results that show that PureConnect did help in enhancing the well-being for environmental justice communities who live in the affected areas. First, we analysed the average feelings of the users in two cohorts. We  found that compared to pre-intervention phase (Cohort 1 and Cohort 2) that was conducted prior to the PureConnect usage (See {\it https://www.sjeqdenver.com/}), the average level of positive feelings was higher in Cohort 3 and Cohort 4, and the average level of negative feelings was lower in Cohort 3 and Cohort 4. This difference was significantly pronounced for the feelings of happiness and loneliness. This indicates that the interventions we introduced did help in improving users' well-being. However, we do note that we introduced three interventions as mentioned in Section~\ref{sec:study-preintervention}, and so the improvement in user feelings is due to the combined impact of three interventions and not just PureConnect alone.

\par Second, to dig deeper and see if PureConnect had some effects on feelings or not, we further investigated the level of usage of PureConnect with users' levels of feelings. We found that there was a positive correlation between the total number of interactions and positive feelings and negative correlation between the total number of interactions and negative feelings. We found this to be significant for all feelings. This  clearly shows that PureConnect usage positively contributed to improving users' well-being.

\par Next, when we conducted sentiment analysis on message shared in PureConnect, and found that positive sentiments were significangtly higher than negative sentiments, even when users discussed problematic issues in the community. Their language was positive, such as when one use shared that they are so excited to make changes in the community, or when they discussed their concerns with each other, they felt like they are not alone in this and have support. 
Finally, our analysis of survey responses clearly show that users found PureConnect very useful.

\par It is important to note that all our analysis is based on correlation and does not establish a causal link between PureConnect usage and people’s wellbeing. Further detailed investigation is needed to establish causality.

%bibliogrpahy
% \bibliographystyle{00}
\bibliographystyle{IEEEtran}
\bstctlcite{MyBSTcontrol}
\bibliography{references}

\end{document}